# Tunneling spin valves based on Fe$_3$GeTe$_2$/hBN/Fe$_3$GeTe$_2$ van der Waals heterostructures

Zhe Wang[*,†], Deepak Sapkota[‡], Takashi Taniguchi[§], Kenji Watanabe[§], David Mandrus[‡,‖,⊥] and Alberto F. Morpurgo[*,†]

[†]DQMP and GAP, University of Geneva, 24 Quai Ernest Ansermet, CH-1211 Geneva, Switzerland

[‡]Department of Physics and Astronomy, University of Tennessee, Knoxville, Tennessee 37996, USA

[‖]Department of Materials Science and Engineering, University of Tennessee, Knoxville, Tennessee 37996, USA

[⊥]Materials Science and Technology Division, Oak Ridge National Laboratory, Oak Ridge, Tennessee 37831, USA

[§]National Institute for Materials Science, 1-1 Namiki, Tsukuba 305-0044, Japan

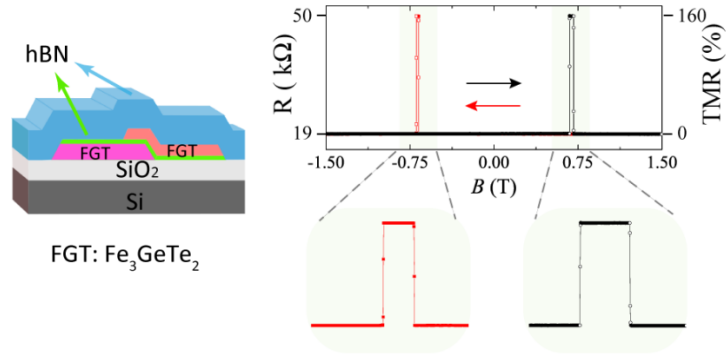

ABSTRACT: Thin van der Waals (vdW) layered magnetic materials disclose the possibility to realize vdW heterostructures with new functionalities. Here we report on the realization and investigation of tunneling spin valves based on van der Waals heterostructures consisting of an atomically thin hBN layer acting as tunnel barrier and two exfoliated Fe$_3$GeTe$_2$ crystals acting as ferromagnetic electrodes. Low-temperature anomalous Hall effect measurements show that thin Fe$_3$GeTe$_2$ crystals are metallic ferromagnets with an easy axis perpendicular to the layers, and a very sharp magnetization switching at magnetic field values that depend slightly on their geometry. In Fe$_3$GeTe$_2$/hBN/Fe$_3$GeTe$_2$ heterostructures, we observe a textbook behavior of the tunneling resistance, which is minimum (maximum) when the magnetization in the two electrodes is parallel (antiparallel) to each other. The magnetoresistance is 160% at low temperature, from which we determine the spin polarization of Fe$_3$GeTe$_2$ to be 0.66, corresponding to 83% and 17% of majority and minority carriers, respectively. The measurements also show that –with increasing temperature– the evolution of the spin polarization extracted from the tunneling magnetoresistance is proportional to the temperature dependence of the magnetization extracted from the analysis of the anomalous Hall conductivity.

This suggests that the magnetic properties of the surface are representative of those of the bulk, as it may be expected for vdW materials.

MAIN TEXT:

In van der Waals (vdW) layered materials, the absence of covalent bonds between the layers is responsible for the very high quality of the material surface that is commonly observed. That is because –contrary to what happens in a covalently bonded solid– at the surface of a vdW layered material no parasitic electronic state originating from broken covalent bond is present. This property has important implications. For instance, it allows atomically thin layers of unprecedented electronic perfection (i.e., so-called 2D materials) to be produced by simple exfoliation of bulk crystals[1, 2]. It also allows controlled heterostructures between 2D materials to be realized by simply stacking different atomically thin crystals on top of each other[2-4]. More in general, owing to the ease with which very high quality surfaces can be produced, the use of vdW materials can facilitate the observation of physical phenomena that are strongly sensitive to the presence of surface defects and disorder of different type.

Here we report on the realization of high-quality tunneling spin valve -conventionally also referred to as magnetic tunnel junction- devices based on nano-fabricated tunnel junctions with vdW ferromagnetic electrodes, and on their use to determine the spin polarization in the vdW ferromagnets themselves. The operation of a spin valve relies on the fact that the current flowing through a tunnel junction is proportional to the product of the density of states (DOS) at the surface of the two metallic electrodes. If the electrodes are ferromagnetic, the difference in the DOS of the majority and minority spins causes the junction resistance to depend on the relative orientation of the magnetization[5-10]. The difference in resistance between the parallel and antiparallel configurations allows the spin polarization in the ferromagnetic electrodes to be extracted quantitatively[5, 10, 11]. In practice, however, a reliable determination of the spin polarization is often not straightforward, because a proper operation of spin-valve devices can be easily prevented by the poor electronic quality of the interface between ferromagnetic electrodes and of the insulating tunnel barrier (see, for instance, Ref [12, 13]) . This may be due, for instance, to surface oxide layers with uncontrolled magnetic properties or defects in the barrier affecting the spin of the tunneling electrons. In vdW ferromagnets these phenomena are not expected to occur and it is interesting to establish whether the anticipated high surface quality can be exploited to realize tunneling spin valves that exhibit electrical characteristics of comparably high quality and enable spin-polarization measurements.

Our experiments rely on $Fe_3GeTe_2$ (see Fig. 1a), a vdW metallic ferromagnetic compound with a relatively high Curie temperature ($T_c \sim 220$ K, not too far below room temperature)[14-29]. $Fe_3GeTe_2$ layers with thickness ranging from approximately 6 to 50 nm and typical linear dimensions of few tens of micrometers can be readily exfoliated with adhesive tape from bulk crystals and used –in conjunction with atomically thin hexagonal boron nitride (hBN) layers– to assemble tunneling spin valve devices as described in detail below. Prior to realizing this type of devices, we investigated the magneto-transport properties of the exfoliated $Fe_3GeTe_2$ crystals themselves, to characterize their ferromagnetic state by measuring the anomalous Hall effect (AHE)[30]. Fig. 1b shows the Hall resistance measured at $T = 4.2$ K on one of these devices (see the microscope image in the left inset of Fig. 1b and the device schematics in the right inset), exhibiting a characteristic behavior representative of what is observed in exfoliated $Fe_3GeTe_2$ thin flakes. A finite anomalous Hall resistance is clearly visible, with a rather sharp transition from positive to negative values at a switching field of approximately -0.7 T accompanied by a clearly defined hysteresis cycle around **B**=0 T upon sweeping the magnetic field from positive to negative values and back (a small asymmetry in the precise value of the switching field for positive and negative applied field is present in most devices, whose origin is unknown). Both the pronounced hysteresis and the sharp switching are worth commenting, because neither of these phenomena is observed in either magnetization or Hall resistance measurements performed on macroscopic bulk crystals[23]. We attribute the difference to the fact that exfoliated $Fe_3GeTe_2$ layers consist of only a small number of domains at low temperature, whereas a very large number of domains are present in macroscopic crystals used for magnetization measurements. Indeed, in atomically thin layers, recent low-temperature optical microscopy studies have shown domain sizes typically larger than 3x3 $\mu m^2$ [26]; in the layers used here, typically 10-50 monolayer thick, the results of our measurements suggest even considerably larger domain sizes. These considerations are relevant for the realization of spin valves based on $Fe_3GeTe_2$ because they indicate in junctions having areas of a few micron squares, mono-domain behavior can be achieved, which is important to facilitate extracting information from the measurements. Finally, note that the observation of a very abrupt switching upon the application of a perpendicular magnetic field –enabled by the presence of a small number of domains– confirms that the easy axis of spontaneous magnetization is perpendicular to the crystalline layers.

To form a tunneling spin valve, two different exfoliated $Fe_3GeTe_2$ crystals are stacked on top of each other, separated by a thin hBN layer (typically 1 or 2 monolayer thick). Fig. 1c shows a schematic representation of such a structure and Fig. 1d an optical microscope image of an actual device. The $Fe_3GeTe_2$ exfoliated layers are produced in a glove box with sub-ppm concentration of oxygen and water and the spin valve is subsequently assembled (on a $Si/SiO_2$ substrate) by means of a commonly used pick-up and transfer technique[31]. The $Fe_3GeTe_2/hBN/Fe_3GeTe_2$ stack is covered with a top hBN layer of typically 30-50 nm thick, to prevent $Fe_3GeTe_2$ from oxidizing upon exposure to air once the device is taken out of the glove box (indeed, $Fe_3GeTe_2$ slowly oxidizes if exposed to ambient conditions for a long time, i.e. for many hours to a day). Chromium/gold contacts to the different $Fe_3GeTe_2$ electrodes are defined by electron beam lithography, electron beam evaporation and lift-off. Prior to deposition of the chromium/gold layers, the top hBN encapsulating layer in the contact region is removed by etching with a $CF_4$ plasma through the same PMMA mask used to lift-off the evaporated metals. After this step, we take care to load devices into the vacuum chamber of the electron-beam evaporator as soon as possible to minimize the exposure of $Fe_3GeTe_2$ to air (in practice, the devices are removed from a reactive ion etcher and immediately loaded into the vacuum chamber of an electron beam evaporator with a few tens of seconds).

The observation of a spin-valve effect in the tunneling resistance requires having two ferromagnetic electrodes with different switching fields[10, 11]. Our devices exploit the fact that the switching field depends slightly on the electrode geometry (i.e., on the geometry of the exfoliated $Fe_3GeTe_2$ layers): albeit the difference is small, it is sufficient for our purposes as we are about to show. Since the magnetization of $Fe_3GeT_2$ points perpendicular to the layers, to demonstrate the occurrence of spin-valve behavior we measure the resistance of the tunnel junction as a function of a perpendicular magnetic field. The result of measurements performed on one of our devices at 4.2 K is shown in Fig. 2 (with reference to the contact scheme shown in Fig. 1(d), measurements are done by sending a 3 nA ac current from electrode $I_{1+}$ to electrode $I_{2+}$ while using electrodes $I_{1-}$ and $I_{2-}$ to measure the voltage drop across the junction with a lock-in amplifier). Upon sweeping *B* from negative to positive values (black circles) an abrupt increase in resistance from approximately 19 to 50 kΩ is observed close to **B** ~ 0.7 T followed by an equally abrupt decrease back to 19 kΩ at a slightly larger magnetic field. As the magnetic field is swept back (red circles), no feature is observed for positive **B**. After reversing the direction of the applied field, however, an analogous increase and a subsequent decrease in tunneling resistance are seen at **B** ~ -0.7 T. This is precisely the behavior expected for a tunneling spin-valve, due to the hysteresis in the switching of the magnetization of the two ferromagnetic electrodes.

To substantiate this conclusion in more detail, we zoom in on the magnetic field interval in which the increase and subsequent decrease in resistance –i.e., the resistance "jumps"– are observed. Data for, respectively, negative and positive values of **B** is plotted in the top panels of Fig. 2b and 2c. The bottom panels of the same figures show the evolution of the anomalous Hall resistance with applied magnetic field that we measured separately on each one of the two ferromagnetic electrodes forming the spin-valve (see Fig. 1d for the configuration of the contacts attached to each of the $Fe_3GeTe_2$ electrodes). The abrupt jumps that are observed in the anomalous Hall resistance mark the precise magnetic field values at which the magnetization of the corresponding $Fe_3GeTe_2$ layer reverses. It is apparent that these values perfectly coincide with the values of **B** at which the spin valve resistance increases abruptly and subsequently decreases back to the original value. This correspondence directly demonstrates that the switching in the resistance of the spin valve is caused by transitions in the relative orientation of the magnetization (i.e., parallel/antiparallel/parallel) that occur upon sweeping the applied magnetic field, as the magnetization in each layer reverses. The extremely sharp switching observed indicates that –at low temperature– the part of the $Fe_3GeTe_2$ electrodes forming the tunnel junction reverse its magnetization as a single domain.

Having established that $Fe_3GeTe_2$/hBN/$Fe_3GeTe_2$ vdW heterostructures do behave as high quality tunneling spin-valves, we now focus on the value of tunneling magnetoresistance (TMR). As it is customary for tunneling spin valves, TMR is defined as $(R_{AP}-R_P)/R_P$, where $R_{AP}$ and $R_P$ represent the resistance measured for parallel and antiparallel alignment of the magnetization in the two ferromagnetic electrodes. The TMR is found to be as large as 160% at 4.2 K. This value is two orders of magnitude larger than that measured in previously studied tunneling spin valves based on conventional ferromagnetic metallic films as electrodes, separated by vdW 2D materials such as graphene[13, 32], hBN[33, 34] and transition metal dichalcogenides[35-37]. The TMR measured in our devices is also ~ 30 times larger than the one that was recently observed in devices formed by exfoliated $Fe_{0.25}TaS_2$ crystals used as electrodes separated by their native oxide acting as tunnel barrier[38].

The measured TMR allows us to further extract the information of spin polarization in $Fe_3GeTe_2$. From Julliere's model the TMR is related to the spin polarization ($P$) in the ferromagnetic electrodes as[5]:

$$TMR = \frac{2P^2}{1-P^2}$$

From the TMR value of 160% it follows that $P = 0.66$ in $Fe_3GeTe_2$, corresponding to having a percentage of majority and minority spins of 83% and 17%, respectively.

Upon warming up the devices, the spin valve behavior persists as the Curie temperature of $Fe_3GeTe_2$ is approached, as shown in Fig. 3a. At higher temperature multiple jumps in the TMR are observed, indicative of switching that occurs with the magnetization reversing for slightly different values of **B** in different part of the junction. Typically, two or at most three steps are seen, implying the reversal of the magnetization direction involves a correspondingly small number of domains during the switching process. To extract the value of the polarization from the TMR we take the steps with the largest resistance. Since –as we just remarked– only two or at most three domains are involved in the magnetization reversal, the configuration corresponding to the largest increase in TMR likely corresponds to the one in which both electrodes are single magnetic domains. This is important for the quantitative determination of the spin polarization $P$ as a function of temperature, which is found to decrease upon increasing temperature (black points in Fig. 3c).

Interestingly, Fig. 3c shows that the extracted spin polarization $P$ exhibits a virtually identical temperature evolution as the anomalous Hall conductivity $\sigma_{xy}$. $\sigma_{xy}$ is calculated from the anomalous Hall resistance (Fig. 3b) and longitudinal resistance (Fig. S1 in supporting information) measured on the same exfoliated $Fe_3GeTe_2$ crystal (Fig. 1b). This may seem surprising because –although both related to the presence of magnetization– spin polarization and AHE probe different physical microscopic processes. In the simplest case, the polarization $P$ is determined by the difference in the density of states for spin up and down at the material surface, whereas the anomalous Hall conductivity is proportional to the bulk magnetization[23]. The constant of proportionality is determined by the longitudinal resistivity $\rho_{xx}$, and also depends on whether the AHE is dominated by extrinsic contributions due to scattering or by the intrinsic properties of the electronic bands[30]. For a generic ferromagnet the longitudinal resistivity has its own (possibly strong) temperature dependence, and therefore the transverse conductivity due to the AHE cannot be expected to scale with temperature in the same way as the magnetization $M(T)$ does.

In $Fe_3GeTe_2$, however, the temperature dependence of the longitudinal resistivity is small, as $\rho_{xx}$ changes only approximately 5% between $T_c = 220$ K and 4.2 K (see Fig. S1), which is why the transverse conductivity $\sigma_{xy}$ exhibits essentially the same temperature dependence of the magnetization. Note also that according to recent work, the AHE in bulk $Fe_3GeTe_2$ is dominated by the intrinsic contribution[23], in which case the proportionality between anomalous Hall conductivity $\sigma_{xy}$ and magnetization upon varying temperature is expected to hold irrespective of the temperature dependence of $\rho_{xx}$. As a confirmation of this conclusion, Fig. 3(c) shows that the measured temperature dependence of the transverse conductivity is perfectly reproduced by the expected functional dependence of the magnetization $M(T) = M(0) \, (1-(T/T_c)^\alpha)^\beta$, a dependence commonly used to interpolate between Bloch's law at low $T$ and the critical dependence of $M(T)$ near $T_c$ [39]. From fitting we find $\beta = 0.35$, very close to the expected 1/3 value and $\alpha = 0.64$ (we have checked that the decrease of $\sigma_{xy}$ upon increasing $T$ is linear in $(T/T_c)^{3/2}$ from the lowest temperature of our measurements up to at least $0.5T_c$). Based on these arguments, our observation that temperature dependence of $P$ and $\sigma_{xy}$ coincide, therefore, provides a clear indication that spin polarization at the surface of $Fe_3GeTe_2$ is proportional to the bulk magnetization. This seems reasonable in a vdW layered ferromagnet such as $Fe_3GeTe_2$, in view of the absence of broken covalent bonds at the surface and of the fact that interlayer couplings are weak compared to intralayer ones.

Finally, we have measured the *I-V* curve and the bias-dependent differential conductance of our spin valves, to probe the evolution of the tunneling magneto-resistance as a function of energy of tunneling electrons. The black and red curve in Fig. 4a represents the *I-V* curves measured at 4.2 K with the magnetization of two ferromagnetic electrodes in parallel and anti-parallel configurations, respectively. In the parallel configuration, the *I-V* curve is nearly perfectly linear, a behavior due to the very large band gap of hBN and to the small thickness of the hBN layer, as a result of which the height and width of the tunnel barrier are virtually independent of the applied bias, and so is the transmission coefficient[40, 41]. Interestingly, however, when the magnetization in the electrodes is in the antiparallel configuration a non-linearity in the *I-V* curve becomes clearly visible. Having just established that the transmission probability through the tunnel barrier is constant as *V* is increased, we conclude that the bias dependence of the *I-V* curve in the antiparallel configuration must originate from the spin of the tunneling electrons.

A possible explanation of the observed behavior is that the non-linearity originates from the energy dependence of the DOS of majority and minority spins in $Fe_3GeTe_2$, since –if transport is determined by elastic tunneling– the differential tunneling conductance is expected to be proportional to the DOS at the energy corresponding to the applied bias. However, this explanation does not seem consistent with the nearly perfect linearity of the *I-V* curves measured when the $Fe_3GeTe_2$ electrodes have parallel magnetization. In addition, a similar steep decrease in TMR upon increasing bias (Fig. 4b for our device) appears to be quite commonly observed in many different types of tunneling spin valves, independently of the ferromagnetic metal used to realize the electrodes[12]. This suggests that the phenomenon does not originate from a specific material property, such as the energy dependence of the DOS. A different possibility is that the application of a large bias opens inelastic tunneling channels causing spin relaxation that concomitantly leads to a suppression in the observed TMR signal. More work will be needed to compare the behavior of $Fe_3GeTe_2$ with that commonly observed in tunneling spin valves with other ferromagnetic electrodes and to reach a definite conclusion.

Irrespective of these details, the results presented here demonstrate that vdW heterosturctures $Fe_3GeTe_2$/hBN/$Fe_3GeTe_2$ exhibit an ideal tunneling spin-valve behavior, which allows us to establish firm conclusions about the magnetic properties of $Fe_3GeTe_2$. The TMR signal is as large as 160% at low temperature corresponding to a spin polarization of 66% at 4.2 K. The observed correspondence between tunneling magnetoresistance and AHE, including the exact correspondence of the magnetic field values at which the two quantities sharply switch and the perfect match of the temperature evolution of both quantities, indicates that the surface spin polarization is proportional to the bulk magnetization. As we have discussed earlier in this paper, the ideality that we have observed in $Fe_3GeTe_2$ is a consequence of the vdW nature of the chemical bonds between the layers. Other vdW ferromagnets are therefore also expected to exhibit an equally ideal behavior, which is why our results should motivate more studies of spintronics devices based on this class of materials, to eventually explore structures comprising exfoliated layers of atomic thickness (i.e., metallic ferromagnetic 2D materials). Indeed, the possibility to combine the type of spintronics functionality that we have demonstrated here with other characteristic properties of atomically thin 2D magnetic materials, such as sensitivity to electrostatic gating, may disclose possibilities unreachable in conventional spintronic devices (think, for instance, of the recently observed dependence of the tunneling probability on the magnetic states of $CrI_3$ tunnel barriers[42-45]). These considerations are particularly timely because the ability to realize individual ferromagnetic $Fe_3GeTe_2$ monolayers[25, 26] with gate tunable magnetic properties has been just demonstrated[25] and because the recent discovery of room temperature ferromagnetism in $VSe_2$ monolayers[46] also suggests that the possibility to use 2D magnetic materials in practical applications is not as unrealistic it may have been believed until now.

**Supporting Information**

Additional data and discussion about longitudinal resistance and anomalous Hall effect of exfoliated $Fe_3GeTe_2$ crystals are available in supporting information.

**AUTHOR INFORMATION**

CORRESPONDING AUTOR

* E-mail: zhe.wang@unige.ch (Z.W.)

* E-mail: Alberto.Morpurgo@unige.ch (A.F.M)

AUTHOR CONTRIBUTIONS

A.F.M. and Z.W. conceived the experiments, D.M. proposed $Fe_3GeTe_2$ as a relevant vdW ferromagnetic metal to use for the ferromagnetic electrodes. Z.W. performed experiments and analyzed the data. D.S. and D.M. grew and characterized the bulk single crystals of $Fe_3GeTe_2$. K.W. and T.T. synthesized hBN crystals. A.F.M. and Z.W. wrote the manuscript with input from all authors.

NOTES

The authors declare no competing financial interest.

ACKNOLEGEMENTS

A.F.M. and Z.W. gratefully acknowledge A. Ferreira and D.-K. Ki for continuous technical support and A. Reddy, H.J. Zhang and Z.P. Wu for experimental help. A.M. and D.M. acknowledge helpful discussions with R. Comin. A.F.M. gratefully acknowledges financial support from the Swiss National Science Foundation, the NCCR QSIT and the EU Graphene Flagship Project. D.S. and D. M. were supported by the Gordon and Betty Moore Foundation's EPiQS Initiative Grant No. GBMF4416. K.W. and T.T. acknowledge support from the Elemental Strategy Initiative conducted by the MEXT, Japan.

FIGURES:

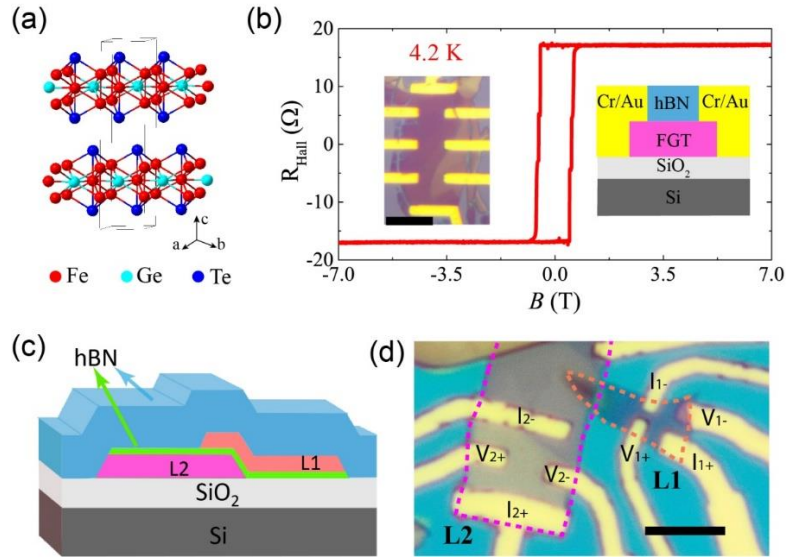

**Figure 1.** Fe$_3$GeTe$_2$: characterization of magnetic properties and device configurations. (a) Crystal structure of Fe$_3$GeTe$_2$. In the ferromagnetic state for T < 220 K the magnetization points along the c-axis. (b) Hall resistance of a 6 nm crystal representative of the typical behavior of the exfoliated layers observed in our studies (data measured with 500 nA ac current at 4.2 K, with *B* applied parallel to the c-axis). The hysteretic behavior around **B** = 0 T, indicative of the presence of a remnant magnetization, is not observed in magnetization measurements on macroscopic crystals that consists of many small magnetic domains. The left and right insets are respectively an optical microscope image of the device used for the measurements (the scale bar is 10 μm) and a schematic of the device cross section. (c) Schematic representation of the van der Waals heterostructures investigated in this work. Two thin Fe$_3$GeTe$_2$ crystals of different shape and thickness (L1 and L2, respectively ~7 and 20 nm thick in the device discussed in the text), are separated by an atomically thin hBN layer. Structures are encapsulated with a thicker hBN layer (typically 30-50 nm thick). (d) Optical microscope image of the actual device used to measure the data presented in the main text. The dotted lines outline the edges of the two Fe$_3$GeTe$_2$ crystals L1 and L2, and the scale bar is 5 μm. The electrodes used as current source (I) and voltage probes (V) in the measurements of the AHE in layer L1 and L2 are indicated in the figure.

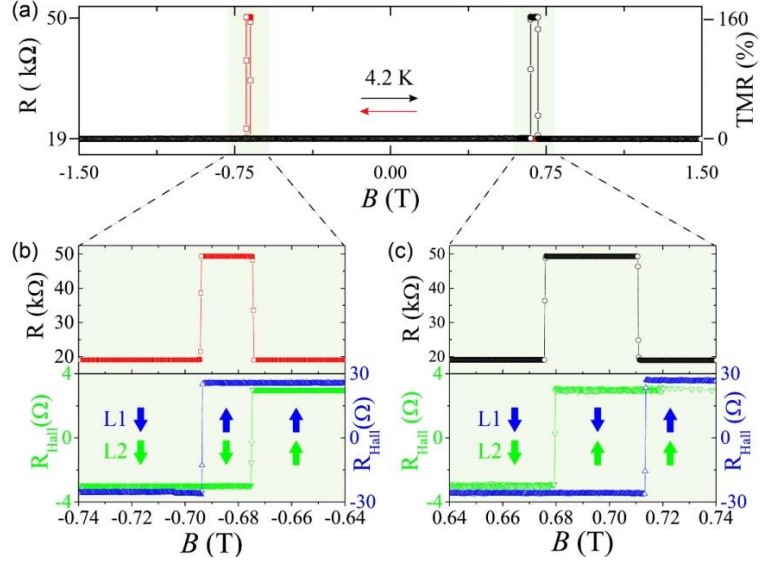

**Figure 2.** Spin valve effect in $Fe_3GeTe_2/hBN/Fe_3GeTe_2$ van der Waals heterostructure. (a) Tunneling resistance of a $Fe_3GeTe_2/hBN/Fe_3GeTe_2$ vdW heterostructure measured at $T$ = 4.2 K with $B$ applied parallel to the $Fe_3GeTe_2$ c-axis. Very sharp resistance jumps are observed for **B** ~ +/- 0.7 T. The variation in tunneling magnetoresistance is ~ 160%. (b-c) Upper panels: zoom-in of the magnetoresistance around +/- 0.7 T. Lower panels: anomalous Hall effect measurements performed on the two $Fe_3GeTe_2$ electrodes forming the heterostructure (i.e., crystals L1 and L2 in Fig. 1 (c) and (d); see Fig. 1(d) for the contact configuration. The Hall resistance is measured with 1 μA and 5 μA ac current for crystals L1 and L2, using in each layer the contacts labeled with I+/- and V+/- to send current and measure voltage). For each electrode, the sharp switches originate from reversal of magnetization, and the blue and green arrows indicate the magnetization orientation at different magnetic field in L1 and L2 determined from the anomalous Hall resistance of the corresponding layer. We establish directly from these measurements that the "jumps" in tunneling magnetoresistance are due to the reversal of the magnetization in each of the electrodes and that the resistance of the tunnel junction is minimum/maximum when the magnetization of two layers is parallel/antiparallel, as expected for a tunneling spin valve.

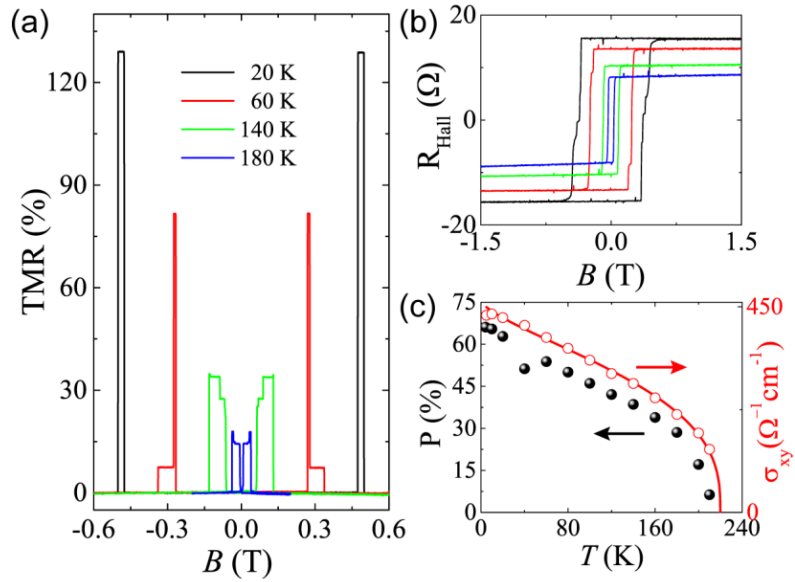

**Figure 3.** Temperature evolution of the tunneling magnetoresistance and of the anomalous Hall resistance. (a) Evolution of the tunneling magneotresistance measured on the device shown in Fig. 1(d) at 20 K (black), 60 K (red), 140 K (green) and 180 K (blue). Below the Curie temperature of $Fe_3GeTe_2$ (~220 K), all magnetoresistance measurements show a clear spin-valve effect. Upon increasing temperature, multiple jumps in the tunneling magnetoresistance become visible, indicating the corresponding formation of multiple magnetic domains in the ferromagnetic electrodes. (b) Hall resistance measured on the device shown in Fig.1b at different temperatures (in panel (a) and (b), lines of the same color correspond to measurements performed at the same temperature). (c) Temperature dependence of the spin polarization extracted from the tunneling magnetoresistance (black filled circles) and of the anomalous Hall conductivity (red open circles). The red continuous line is a fit of $\sigma_{xy}$ with the functional form commonly used to fit the with the temperature dependence of the magnetization $M(T) = M(0) (1-(T/T_c)^{\alpha})^{\beta}$, with $\alpha = 0.64$ and $\beta = 0.35$. The virtually identical evolution upon increasing $T$ indicates that the magnetic properties of the surface (probed by the tunneling magnetoresistance) are representative of the bulk magnetic properties (probed by the anomalous Hall effect).

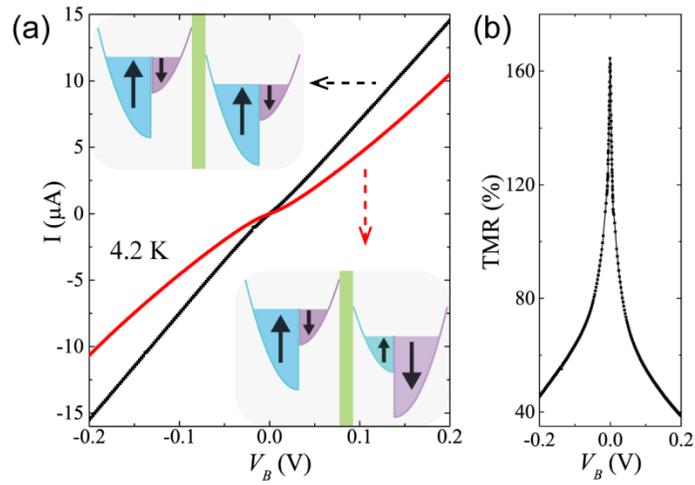

**Figure 4.** Bias dependence of the tunneling magnetoresistance. (a) *I-V* curves measured at *T* = 4.2 K with the magnetization in the two $Fe_3GeTe_2$ electrodes pointing parallel (black curve, *B* = 0 T) and anti-parallel (red curve, **B** = -0.68 T) to each other. The insets show the corresponding configuration in density of states of majority and minority spins in the two electrodes. (b) Bias dependence of tunneling magnetoresistance at 4.2 K. A steep decrease is observed already upon the application of a small dc bias.

# Supporting Information

**Temperature dependence of the longitudinal resistance.** In the main text we discussed that the magnitude of the anomalous Hall effect (AHE) is proportional to the magnetization. We also indicated that the constant of proportionality depends on the longitudinal resistivity in a way that depends on the microscopic mechanism responsible for the AHE (of course the details depend on whether one discussed the anomalous Hall resistivity of conductivity). As explained in the main text, this is important because the longitudinal resistivity of a ferromagnet can depend strongly on temperature. If so, care is needed in relating the temperature dependence of the AHE to that of the magnetization.

In Fig. S1 we show the longitudinal resistivity measured on the same device whose transverse conductivity data are shown in Fig. 3(c) of the main text. Throughout the entire temperature range below $T_c$, the longitudinal resistivity changes by only a few percent, i.e. it is essentially constant. This implies that –within a very good approximation– the temperature dependence of the AHE (either conductivity or resistivity) is proportional to the temperature dependence of the bulk magnetization $M(T)$.

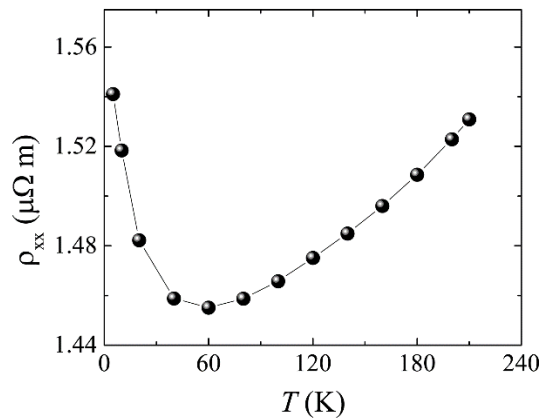

**Figure S1**. Temperature dependence of longitudinal resistivity measured on the device shown in Fig.1b in main text, whose $\sigma_{xy}$ data are shown in Fig. 3(c).

**Anomalous Hall resistance of exfoliated $Fe_3GeTe_2$ crytsals.** In the main text we have argued that even in rather large exfoliated crystals of $Fe_3GeTe_2$ (with linear dimensions of ~ 30 μm) magnetization reversal near the switching field occurs through the formation of only a small number of domains. To illustrate this fact, Fig. S2 zooms in on the data shown in Fig. 1(b) of the main text, measured on one of these large exfoliated crystals. It is apparent that the reversal of the anomalous Hall resistance occurs in three sharp steps, indicative of the presence –near the switching field– of as many magnetic domains.

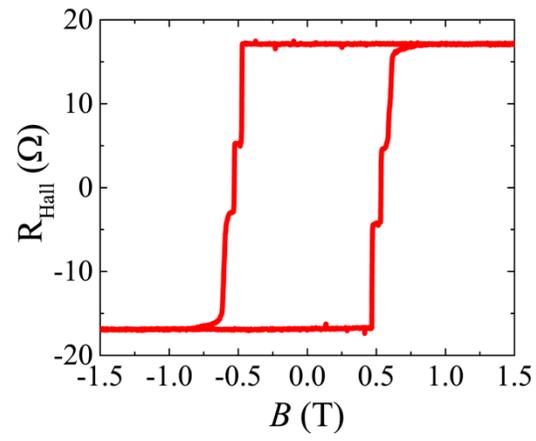

**Figure S2**. Zoom-in part of anomalous Hall resistance shown in Fig. 1b.